\documentstyle[aps,epsfig,twocolumn]{revtex}
\begin{document}
\draft
\flushbottom
\twocolumn[
\hsize\textwidth\columnwidth\hsize\csname @twocolumnfalse\endcsname

\title{
Surface electronic structure and magnetic properties of 
doped manganites}
\author{M. J. Calder\'on$^{1,2}$, L. Brey$^1$ and F. Guinea$^1$}
\address{
$^1$ 
Instituto de Ciencia de Materiales (CSIC). Cantoblanco,
28049 Madrid. Spain. \\
$^2$ Departamento de F{\'\i}sica Te\'orica de la Materia Condensada,
Universidad Aut\'onoma de Madrid, 28049 Madrid. Spain. \\}
\date{\today}
\maketitle 
\tightenlines
\widetext
\advance\leftskip by 57pt
\advance\rightskip by 57pt

\begin{abstract}
The electronic structure and magnetic properties of 
La$_{1-x}$A$_{x}$MnO$_3$ are investigated. It is assumed that,
at the outermost layer, the environment of the Mn ions does not have 
cubic symmetry. The $e_g$ orbitals are split and the double exchange
mechanism is weakened. The charge state of the Mn ions is modified,
and the magnetic ordering of the spins tends to be antiferromagnetic.
The surface magnetization and 
the  dependence of the transport properties through
the resulting surface barrier on applied
magnetic field and temperature is analyzed.
\end{abstract}

]
\narrowtext
\tightenlines
Doped manganites show many unusual features, the most striking being
the colossal magnetoresistance 
in the fully ferromagnetic phase~\cite{WK55,CVM97}.
Extensive research shows that the transport properties, and the
magnetoresistance in particular, are significantly modified
at artificially created  
barriers~\cite{HCOB96,Letal96,Getal96,Metal97,Vetal97,SARP98,Setal98,SARE98}
or in ceramic materials~\cite{Setal96,Metal96,BFMO97,FMBO97,BFMO98}.
The magnetoresistance, as function of temperature, drops more
rapidly than in the bulk. It has larger values at low fields, 
and persists at large fields, unlike in the bulk case.
The relevance of the interfaces
in perovskite manganites can also be inferred
by comparing with transport in related materials
which exhibit colossal magnetoresistance~\cite{HC97}.
These properties have been ascribed to changes in the
interface structure, although the origin of these modifications,
and the resulting structure are not known.

In the present paper, we analyze the simplest, and most common,
modification with respect to the bulk 
that a surface may show:
the loss of cubic symmetry around the Mn ions. It is well known
that La$_{1-x}$A$_x$MnO$_3$ shows a transition from a
tetragonal (or orthorhombic) structure to a cubic one as the
value of the doping $x$ is increased. The systems with the
highest Curie temperature have $x \sim \frac{1}{3}$ and are in the
cubic phase. This implies that the two $e_g$ orbitals of the
Mn ion are degenerate and contribute to the conduction band.
In this situation, 
the double exchange mechanism is enhanced.

The cubic symmetry is lost at the surface. If the last
layer is oxygen deficient, the resulting splitting between 
the $e_g$ orbitals can be larger than typical Jahn-Teller
splittings in La$_{1-x}$A$_x$MnO$_3$ with small values of $x$.
When one of the $e_g$ orbitals moves away from the Fermi level,
the double exchange mechanism is weakened, and direct
antiferromagnetic couplings between the core $S=\frac{3}{2}$ spins
can prevail. Moreover, the reduction in electronic kinetic energy
can also lead to charge transfer between the surface layers
and the bulk, contributing to the formation of a surface dipole.
All these effects can be modified by surface spin waves,
which, in turn, depend on temperature and external magnetic
fields. A significant dependence of the metal-insulator transition
temperature as function of oxygen pressure in thin films is
found in~\cite{Retal98}.

In order to investigate these features, we start from a tight
binding Hamiltonian, using the two $e_g$ orbitals, $d_{x^2-y^2}$
and $d_{3z^2-r^2}$, which we  designate $x$ and $z$ respectively. 
Hopping between them takes place through virtual jumps to
the intermediate O ions. Fixing the orientations of the $e_g$
orbitals to the frame of reference of the lattice, we obtain for the
$z-$ direction, $t_{zz} = t$ and $ t_{xz} = t_{xx} = 0$. For the
directions in the $x-y$ plane $t_{zz} = \frac{1}{4} t , t_{xz}
= \pm \frac{\sqrt{3}}{4} t$ and $t_{xx} = \frac{3}{4} t$, where the
two signs in $t_{xz}$ correspond to the $x$ and $y$ directions,
and $t$ is the effective $e_g - e_g$ hopping
generated from the $(dp\sigma )$ matrix element between a $d$ orbital
in a given Mn ion, and a $p$ orbital in a neighboring O ion\cite{PP98,feinberg}.
At each site there is also a spin, from the three $t_{2g}$ orbitals,
which we treat as classical, and parametrize in terms of the
Euler angles $\theta$ and $\phi$.
We assume that the Hund's coupling between the $e_g$ electrons and this
spin is much larger than other scales in the model. Then, the hopping elements
depend on the orientation of the core spins, and the actual hopping
is\cite{muller}:
\begin{equation}
t_{\alpha , \beta}^{i,j} = t_{\alpha , \beta} \left(
\cos \frac{\theta_i}{2} \cos \frac{\theta_j}{2} +
\sin \frac{\theta_i}{2} \sin \frac{\theta_j}{2} e^{i ( \phi_i 
- \phi_j )} \right)
\label{hopping}
\end{equation}
where the value of the hopping has been estimated $t \sim 0.1 - 0.3$eV.

We study a cubic lattice with periodic boundary conditions in the
$x$ and $y$ directions, and open boundary conditions along the
$z$ direction, where two (001) surfaces terminate the lattice. 
We use slabs with a thickness of 20 atoms. 
We have checked that, for this size, the bulk properties are 
recovered at the center of the slab.
The environment of the outermost Mn ions is deficient in oxygen.
The oxygen octahedra which surround the Mn ions are incomplete.
The absence of the negatively charged O$^{2-}$ ions leads to
a downward shift of the $e_g$ levels with respect to the values
in the bulk. This shift is larger for the $d_{3z^2-r^2}$ orbital,
which points towards the surface. The $d_{x^2-y^2}$ orbital
is more localized around the ion, and is less
sensitive to the change in the environment. In order to keep the number of
free parameters in the model to a minimum, we leave the
$d_{x^2-y^2}$ level unchanged with respect to the bulk, 
while the $d_{3z^2-r^2}$ is shifted downwards by an
amount $\Delta$.
The value of $\Delta$ should be comparable, or larger, than the
observed Jahn-Teller splitting in LaMnO$_3$, which, in turn,
is larger than the $e_g$ bandwidth. Reasonable values
of $\Delta$ are $\sim 0.5 - 1.5$eV\cite{CVM97}.
The $e_g$ levels at all other layers remain unchanged, except
for electrostatic effects (see below).

This shift of the surface $d_{3z^2-r^2}$ orbitals leads to charge transfer
between the bulk and the surface. We treat the induced electrostatic
effects within the Hartree approximation. This gives rise
to an additional shift in the electronic levels which is determined by solving 
selfconsistently the Schr\"odinger and the Poisson equations.
This electrostatic shift is equal for the two $e_g$ orbitals
at each layer. 
Screening from other levels is 
described in terms of a dielectric constant of value $\epsilon
= 5 \epsilon_0$. Selfconsistency is imposed on the $e_g$ levels in
all layers in the system. We do not consider the possibility
of orbital ordering induced by a Hubbard $U$ between electrons
at the two $e_g$ orbitals\cite{KS97}.
We analyze underdoped materials, $x < 0.5$, where no
unusual magnetic textures are expected\cite{BK98}.
The Hartree approximation acts to suppress charge fluctuations, although
it does not split the $e_g$ bands in the way a Hubbard term does.
This effect should be less important at the surface, due to the
crystal field splitting $\Delta$ introduced before.

\begin{figure}
\centerline{\epsfig{file=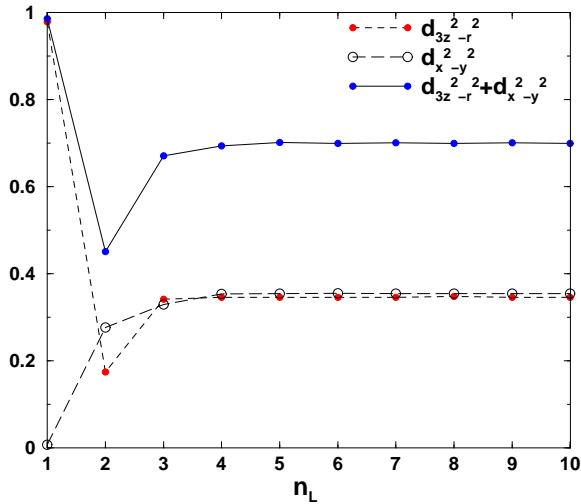,width=3in}}
\caption{Charge in the $e_g$ orbitals as a
function of the distance from the surface for $\Delta = 10 t$.
On the surface, $n_L = 1$, the $x$ orbitals are almost empty, 
due to the shift $\Delta$.}
\label{fig:charge}
\end{figure}

Typical results, for $\Delta = 10 t$,  hole concentration of 
$x$=0.3 and a ferromagnetic 
configuration of all spins are shown in fig.~[\ref{fig:charge}].
Charge neutrality corresponds to a total occupancy of the $e_g$
orbitals of $0.7$.

The large splitting between the $e_g$ orbitals at the surface
leads  to a significant charge transfer to the outermost layer.
The charge distribution reaches the bulk values at the third layer,
in agreement with the expected short screening length of the metal.
The surface $d_{3z^2-r^2}$ orbitals are almost full, while the  $d_{x^2-y^2}$
are empty, so that the charge state of the ion is Mn$^{3+}$.
Hence, for the value of $\Delta$ used in fig.~[\ref{fig:charge}], the double 
exchange mechanism is almost completely suppressed at the surface.
%
%
Higher values of $\Delta /t$ do not
alter these results.
For intermediate values of $\Delta  /
 t $, the surface electrostatic barrier depends significantly
on the magnetic configuration.
The surface
$d_{x^2-y^2}$ orbitals are practically empty, and the  $d_{3z^2-r^2}$ is
occuppied for $\Delta > 3 t$. 

\begin{figure}
\centerline{\epsfig{file=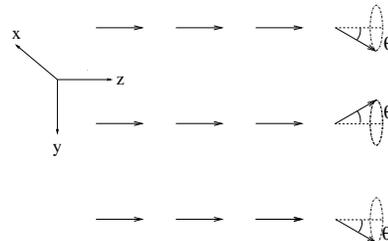,width=2in}}
\caption{Magnetic structure considered in the text. Only the spins
at the surface layer are rotated, in the way shown.}
\label{fig:spins}
\end{figure}

The suppression of the double exchange ferromagnetic 
coupling at the surface leads
to an enhancement of the superexchange interaction among the Mn ions. 
To investigate further this effect, we study magnetic 
configurations where the core spins of the surface Mn ions
are allowed to rotate, as shown in fig.~[\ref{fig:spins}].
The canting angle $\theta$ is used as a variational parameter.
$\theta = 0$ implies a perfect ferromagnetic order at the surface.
$\theta = \pi / 2$ gives rise to an antiferromagnetic alignment
of the surface spins, at right angles with the bulk magnetization.
To take into account the superexchange interaction, 
we introduce an antiferromagnetic coupling between the Mn core spins.
Substracting a trivial constant, the energy per surface ion, due to
this coupling, is:
\begin{equation}
E_{AF} = J_{\perp} \cos ( \theta ) + 2 J_{\parallel} \cos ( 2 \theta )
\label{energyaf}
\end{equation}
where $\theta$ is the angle shown in fig.~[\ref{fig:spins}], $J_{\perp}$
is the antiferromagnetic coupling between a Mn spin at the surface and
one in the next layer, and $J_{\parallel}$ is the coupling between
spins at the surface layer. From symmetry considerations,
$J_{\parallel} = 4 J_{\perp}$\cite{M97}.


%

The total energy of the system is the sum of the kinetic, 
Hartree and magnetic
energies. By minimizing the total energy 
as a function of $\theta$ we obtain the canting angle 
as a function of the antiferromagnetic coupling $J_ {\perp}$.
 
\begin{figure}
\centerline{\epsfig{file=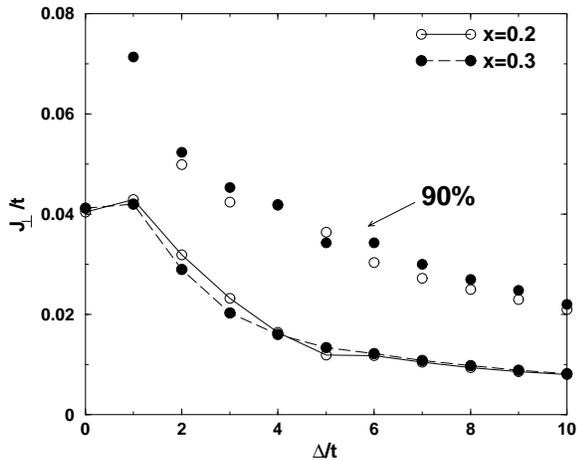,width=3in}}
\caption{Magnetic phase diagram of the surface of a doped manganite.
The lines separate the fully ferromagnetic and the canted regions.
The dots indicate that the order is $90\%$  antiferromagnetic, namely,
$\theta=81^o$.
Full dots are results for hole concentration of $x=0.3$. Open dots
are for $x=0.2$.}
\label{fig:phasediagram}
\end{figure}

Fig.~[\ref{fig:phasediagram}] shows the calculated phase diagram as function
of the splitting at the surface between the $e_g$ levels, $\Delta$, and
the direct antiferromagnetic coupling between the core spins $J_{\perp}$,
for two different values of the hole concentration.

Realistic values of $J_{\perp}$ are of the order of 0.02$t$\cite{perring}
and from fig.~[\ref{fig:phasediagram}], we see that,
for this value, the value of $\theta$ is close to that
of perfect surface antiferromagnetism, $\pi / 2$. Our approach
tends to underestimate this tendency towards antiferromagnetism,
as we do not allow to relax the spins in the layers
deeper into the surface. 

\begin{figure}
\centerline{\epsfig{file=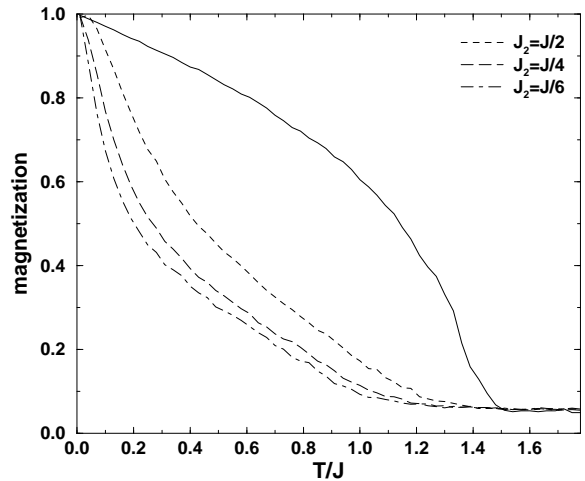,width=3in}}
\caption{Magnetization of the last layer
as function of temperature for a cluster with 20
layers and open boundary conditions. For comparison, the results for
a cluster of the same size and periodic boundary conditions (no
surfaces) are also shown.}
\label{fig:surfacemag}
\end{figure}

The changes in the magnetic surface structure also lead to modifications
in the spin stiffness at the surface, which is weaker than in the
ferromagnetic bulk. It is straightforward to show that weaker couplings
lead to the formation of bands of surface magnons\cite{M88}. We find such
bands both in the canted and in the ferromagnetic phases shown
in fig.~[\ref{fig:phasediagram}]. The bandwidth is narrower than
the bulk magnon band.
Thermal excitation of these modes leads
to a decay of the surface magnetization as function of temperature
which is faster than in the bulk. 
In order to estimate this effects, we have calculated, using 
Monte Carlo techniques\cite{mjc1},
the surface magnetization of a cluster of classical spins. 
In this model, the bulk
double exchange mechanism is described by an 
effective ferromagnetic Heisenberg coupling $J$,
and the surface spins interact with an antiferromagnetic 
coupling $J_1$. The outermost spins interact
with the spins in the second layers
via a ferromagnetic 
coupling $J_2$. From the results reported previously, we 
estimate that $J_1$ is very small, 
$\sim -J/100$ and $J_2$ takes  a value between $J/2$ and $J/6$.
In fig.~[\ref{fig:surfacemag}] we plot the surface 
magnetization as a function of the temperature for
different values of $J_2$. The results 
obtained do not depend strongly on the value of $J_2$ and they
are in reasonable agreement with the 
experimental results reported in\cite{Petal98}.

The combination of energy shifts and changes in the magnetic couplings
reduces the transmission of the surface, even in the absence of
other, extrinsic, barriers. In order to estimate this effect, we have 
calculated the resistance between 
two perfect double exchange ferromagnetic metals
separated by a layer with the amount of charge and the magnetic structure
obtained in  the previous calculations. 
Since only the $d_{3z^2-r^2}$ orbitals contribute 
to the transport through the interface, 
we have simplified the model, and only  a single orbital 
per site is kept. The calculations were done using the 
method described in~\cite{CVB98,V98}.
Two effects contribute to the resistivity, 
the shift in energy of the interface orbital and the 
difference in spin orientation
between the atoms at the bulk and at the interface. 
The conductance is sharply reduced when the surface 
layer is antiferromagnetic,
as the double exchange mechanism is suppressed. 
We find that for hole concentrations of
$x$=0.3 the presence of the interface increases the resistance 
of the system in a factor bigger than 10.

\begin{figure}
\centerline{\epsfig{file=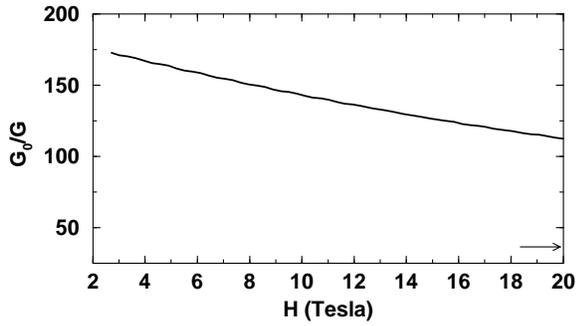,width=3in}}
\caption{Inverse conductance, normalized to the conductance of a perfect
ferromagnetic system ($\theta=0$ and $\Delta=0$), of a magnetic
layer, as function of applied field. The arrow indicates the saturation
limit at high fields.
The size of the system is $20\times20\times20$.
 The parameters used are described in the
text.}
\label{fig:MR}
\end{figure}

A magnetic field
will reduce the antiferromagnetism at the surface, and, at high fields,
the surface spins are aligned parallel to the bulk. We estimate this effect
by adding a magnetic field to the model and 
finding the magnetic structure which 
minimizes the energy. The resulting conductance, for $\Delta = 3 t$
and $J_{\perp} = 0.025 t$, is plotted in fig.~[\ref{fig:MR}]. A very high
field $\sim 60$T is required to saturate the magnetoresistance.
Note that the low field  ( < 1T ) magnetoresistance is probably due to
the alignment of the polarization of the bulk electrodes, which
can be understood within conventional models\cite{J75,S89}.
Our results for the high field dependence are consistent with the available 
experimental data
data~\cite{HCOB96,Letal96,Getal96,Metal97,Vetal97,SARP98,Setal98,SARE98}.

The magnetic excitations of the surface layer also modify the 
magnetoresistance at finite temperatures~\cite{B97,G98,MNV98}. Spin flip
scattering due to thermally excited magnons leads to a
suppression of the magnetoresistance at temperatures below
the bulk Curie temperature. The dependence of the
magnetoresistance on temperature should be similar to that of
the surface magnetization, shown in fig.~[\ref{fig:surfacemag}].

In conclusion, we have shown that the lack of cubic symmetry at surfaces,
combined with the double exchange mechanism, leads to significant
changes in the magnetic and transport properties of doped manganites.
Charge is transferred from the bulk to the surface layers, and 
an antiferromagnetic ordering of the surface spins is favored.
This structure is consistent with the observed
reduction of the magnetoresistance
in junctions and ceramic materials at temperatures below the
bulk Curie temperature,  with the high field dependence
of the magnetoresistance and with the experimental results 
on surface magnetization.

We acknowledge fruitful discussions with A. de Andr\'es,
M. Coey, M. Hern\'andez, J. L. Mart{\'\i}nez, J. Fontcuberta, X. Obradors
and J. A. Verg\'es. This research is supported through grants
PB96/0875 and  PB96/0085. LB amd MJC also acknowledge finantial 
support from the 
Fundaci\'on Ram\'on Areces.   

\end{document}